# Persistence of multifractal dynamics at the mammalian neuromuscular junction across extracellular calcium conditions

A. J. da Silva[1,*] ● S. Floquet[2] ● R.F. Lima[3]

**Abstract** The neuromuscular junction represents a relevant experimental model for revealing important biophysical mechanisms underlying synaptic transmission. In this work, we revisited the time-series analysis of electrophysiological recordings obtained from mammalian diaphragm neuromuscular junctions. We applied multifractal detrended fluctuation analysis (MFDFA) to assess whether intervals between spontaneous potentials exhibit multifractal behavior at physiological extracellular calcium concentrations and at hypocalcemic and hypercalcemic levels. Surprisingly, we observed a robust and stable multifractal profile across all electrophysiological recordings, regardless of calcium concentration. Therefore, our results provide support for a rich and complex scenario governing neurotransmitter release. Based on these findings, we discuss their biophysical significance and physiological implications.

*corresponding author: adjesbr@ufsb.edu.br

[1] Centro de Formação em Ciências Agroflorestais, Universidade Federal do Sul da Bahia, Itabuna, Bahia, 45613-204, Brazil.

[2] Colegiado de Engenharia Civil, Universidade Federal do Vale do São Francisco, 48902-300, Juazeiro, Bahia, Brazil.

[3] Departamento de Fisiologia e Farmacologia, Faculdade de Medicina, Universidade Federal do Ceará, 60430-270, Fortaleza, Ceará, Brazil.

## 1 Introduction

The neuromuscular junction (NMJ) is a specialized synapse that transmits information from the nervous system to muscle [1]. Studying this region has been crucial to understanding the biophysical mechanisms underlying neurotransmitter release at nerve terminals. Using amphibian junctions, Bernard Katz and colleagues began remarkable work with these tissues, observing small depolarizations of the postsynaptic membrane, known as miniature end-plate potentials (MEPPs) [2]. Shortly after the MEPP discovery, Palade identified vesicles in the nerve terminal, providing a plausible physiological substrate for Katz to address the challenge of interpreting the origin of spontaneous electrical discharge [3]. According to his visionary suggestion, the fusion of a single vesicle (also called a quantum) with the nerve-ending membrane was the main event responsible for producing a MEPP. This interpretation is central to the vesicular hypothesis, a view widely accepted in the neurophysiology community [4]. However, researchers needed to examine the cellular and molecular aspects of vesicular fusion, combining different approaches. They systematically identified, modeled, and characterized the exocytosis machinery and the proteins involved in neurotransmission [5]. Within this scientific program, investigating the roles of ions emerged as a key element in explaining specific processes, revealing many aspects of how neurons work [6].

In the central nervous system, ions play important physiological roles, including generating and establishing the resting potential and nerve impulse, facilitating nerve transmission, and participating in various intracellular mechanisms [7]. Consequently, certain ionic species help to maintain homeostasis and allostasis in the brain, and imbalances in their concentrations may contribute to the onset of various diseases [8]. Moreover, it is important to note that there are ions, such as mercury ($Hg^{2+}$) and lead ($Pb^{2+}$), which are extremely toxic and may trigger serious neurological disorders [9]. Beyond that, elevated extracellular potassium levels in the brain are associated with epileptic seizures. In contrast, alterations in extracellular calcium ions ($[Ca^{2+}]_o$) may produce hypo- and hypercalcemia that are related to disturbances such as neuromuscular irritability and muscle cramps, paresthesia, and psychosis [10,11]. In the heart, $Ca^{2+}$ levels below or above the physiological range can cause cardiac arrhythmias and other pathologies [12]. Given their immense physiological importance, $Ca^{2+}$ has attracted significant attention because of its wide-ranging activity, particularly in mechanisms fundamental to nervous system function.

Investigations into how ions modulate the electrophysiological signals produced at the nerve terminal have also advanced our understanding of many biophysical aspects of the NMJ. In this framework, MEPP recordings proved powerful, as they allowed detailed studies of how distinct ions shape neurotransmission. In this sense, the physiological role of $Ca^{2+}$ has reached a new level, underscoring its importance in neuronal communication [13]. Binding of $Ca^{2+}$ to the synaptotagmin site is now widely accepted to trigger vesicle fusion at the synaptic terminal [14]. This mechanism holds in both the NMJ and vertebrate brain synapses, as similarly shown in other synapses from invertebrate species [15]. Moreover, combining electrophysiology and molecular biology clearly shows that $Ca^{2+}$ is a fundamental agent in neuroplasticity mechanisms [16].

Within the scientific program to elucidate the physiological origin and functional basis of the MEPP, Katz and his colleagues, inspired by the MEPP time series, formulated the first mathematical pillar to explain the distribution of amplitudes and intervals of these signals [17]. According to these investigators, the amplitude was satisfactorily described by a Gaussian function, while the intervals between events were explained within a framework based on the Poisson distribution. This statistical description has been tested, yielding different interpretations when compared with Katz's original conclusions. Over the years, researchers have systematically revisited his model of vesicular fusion, essentially interpreted as a random physiological phenomenon. For instance, van der Kloot and colleagues modeled MEPP interval data from amphibian NMJs, accounting for extracellular ionic modifications and finding deviations from Poisson model predictions [18-20].

Several rigorous methods exist today for testing fractality in biological time series. Among these methods, MFDFA, developed by Kantelhardt *et al.*, is notable for uncovering possible scale-correlation properties of the time series [21]. The procedure expands the well-known detrended Fluctuation Analysis (DFA), previously developed by Peng *et al.* to quantify monofractality or long-range correlations in datasets [22]. In the MFDFA case, evaluating the scaling behavior of fluctuations for a range of statistical moments allows the identification of multifractal behavior that can not be obtained by a single exponent provided by the DFA procedure. In neuroscience, MFDFA has been widely applied for different goals, from theoretical investigations to data measured by electrophysiological techniques such as electroencephalography and electrocardiography [23-25].

Regarding the NMJ, Rana and Dey applied DFA to electromyography (EMG) and force-sensitive resistor (FSR) signals to study neuromuscular coordination and postural control [26]. Using MFDFA, Gaurav *et al.* quantified EEG-based cognitive tasks, while Reyes-Manzano *et al.* used ECG data collected from the human heart to compare healthy controls and fibromyalgia patients and found decreased nonlinearity and stronger anticorrelations [27,28]. For the nervous system, Yang *et al.* used MFDFA to estimate long-range memory behavior in neuronal spiking data [29]. Zorick and Mandelkern used the method to investigate patterns in human EEG signals recorded from waking and different sleep stages [30]. These scholars highlighted the importance of MFDFA for automatically classifying different states of consciousness. However, applications were also successful in identifying changes in cortical dynamics in conditions such as parkinsonism and schizophrenia. Thus, MFDFA offers an essential tool to reveal important aspects of physiological dynamics that are not captured by other approaches.

In recent years, Silva and collaborators have focused on understanding the ionic influence on biological systems across different theoretical contexts [31,32]. In particular, these researchers have combined experimental and theoretical approaches to understand whether MEPPs are physiological events generated randomly over time or are governed by long-range correlations. Manipulation of $[Ca^{2+}]_o$ and extracellular potassium concentrations in Ringer suggests that MEPP amplitudes are better described by non-extensive statistics. At the same time, their intervals followed Tsallis statistics and Benford's law [33,34]. Therefore, this subject remains an important research topic within biostatistics and the biophysics of neurotransmission. Identifying patterns in MEPP distribution over time can inform models and therapeutic strategies for evaluating the NMJ in both healthy and diseased conditions.

Combining rigorous electrophysiological data collection with MFDFA allows one to address a key question: how robust is multifractality in neurotransmitter release? In other words, we wish to test whether the multifractal behavior identified under physiological conditions remains stable or changes when $[Ca^{2+}]_o$ falls within hypocalcemic or hypercalcemic ranges, using MEPP time series.

# 2 Materials and Methods

## 2.1 Experimental procedures

The empirical procedures follow the standard protocols established by our previous study [35]. The experimental protocols were approved by the Animal Research Committee (CEUA - UFC, protocol 130/2017). We used male wild-type and Swiss mice aged 70-126 days. Animals were euthanized by cervical dislocation and diaphragms were rapidly extracted and inserted into a physiological artificial fluid (Ringer solution). The following successive $[Ca^{2+}]_o$ were used (mM): 0.6, 1.2, 1.8, 4.8, 10, and 15. The external composition for $[Ca^{2+}]_o$ = (0.6–4.8) mM contained the following concentrations (mM): NaCl (137), $NaHCO_3$ (26), KCl (5), $NaH_2PO_4$ (1.2), glucose (10), and $MgCl_2$ (1.3). In addition, the external solution for $[Ca^{2+}]_o$ = (10–15) mM contained (mM): NaCl (137), $NaHCO_3$ (12), KCl (5), $NaH_2PO_4$ (0.3), glucose (10), and $MgCl_2$ (1.3). In both solutions, pH was adjusted to 7.4 after gassing with 95% $O_2$−5% $CO_2$. We left tissues in the solution for 30 min before electrophysiological recordings began to minimize the mechanical trauma of extraction. Next, we gently transferred the tissues to a recording chamber continuously bathed in artificial solution at T = 24 ± 1°C. A standard intracellular recording technique was performed to monitor the frequency of spontaneous MEPPs by inserting a micropipette into the chosen muscle fiber. We used borosilicate glass microelectrodes with resistances of 10–25 MΩ, filled with 3 M KCl, for electrophysiological recordings. Strathclyde Electrophysiology Software (John Dempster, University of Strathclyde), Origin (OriginLab, Northampton, MA), and MATLAB (The MathWorks, Inc., Natick, MA) were employed for electrophysiological acquisition and data analysis. Fig. 1 shows an electrophysiological segment collected at physiological $[Ca^{2+}]_o$, attesting to the quality of the electrical recordings.

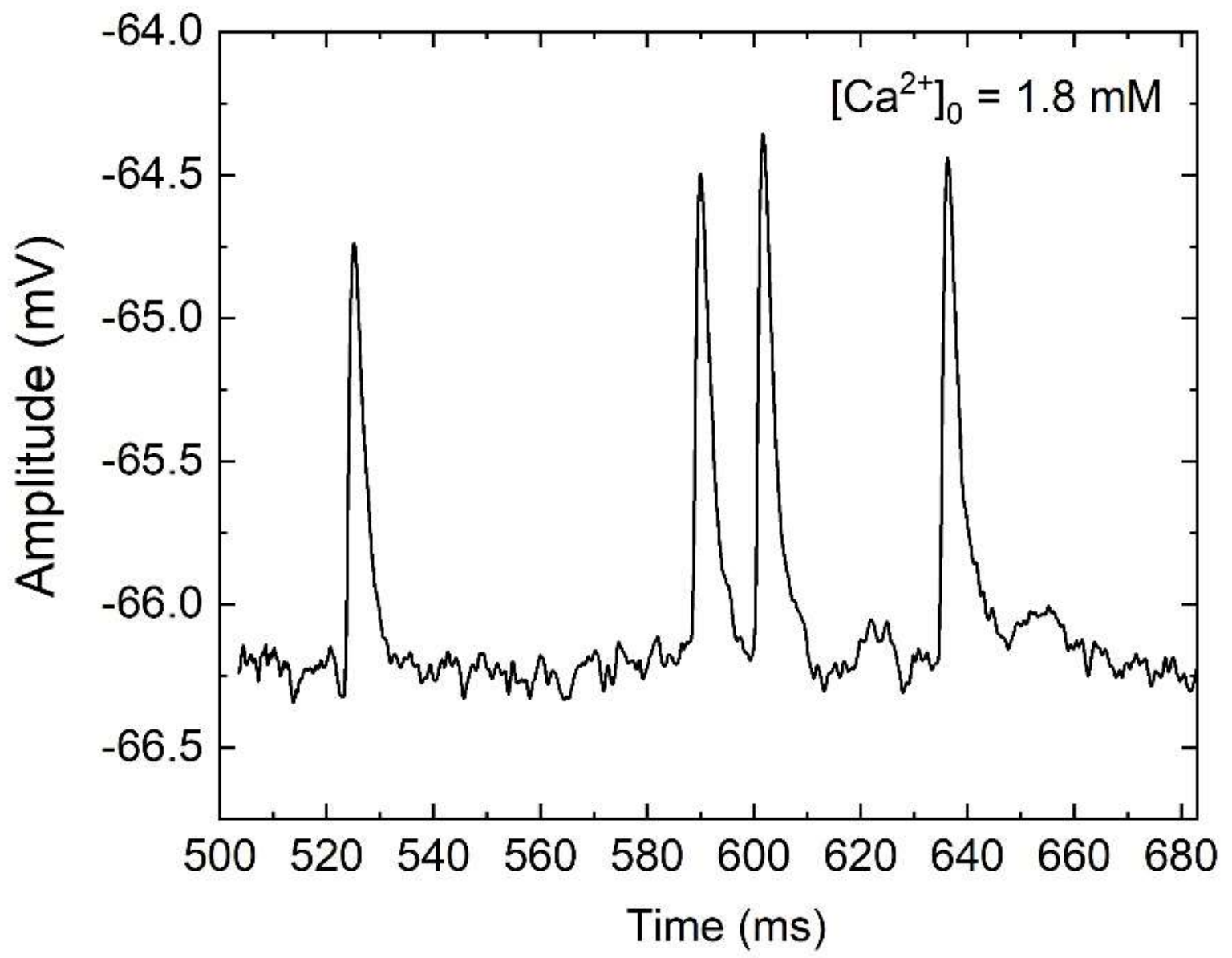

**Fig. 1** A representative interval from an electrophysiological recording

## 2.2 Multifractal Detrended Fluctuation Analysis

We used the approach introduced by Kantelhardt *et al.* to inspect the scaling structure and multifractal properties of the MEPP time series [21]. By considering a range of statistical moments, this method provides a framework for identifying multifractal properties in nonstationary time series. Below is presented a theoretical summary of the MFDFA as originally described in seminal work.

Initially, let us consider a time series $x(i)$, with $i = 1, 2, ..., N,$ where $N$ is the total number of observations. The mean of the series is calculated as:

$$\bar{x} = \frac{1}{N}\sum_{i=1}^{N} x(i) \tag{1}$$

Next, the integrated profile of the series is constructed by calculating the cumulative sum of the deviations from the mean:

$$Y(k) = \sum_{i=1}^{k} \left[x(i) - \bar{x}\right], \; k = 1, \ldots, N \tag{2}$$

For each scale $s$, the profile $Y(k)$ is divided into $Ns = int\ (N/s)$ non-overlapping segments of length $s$. Since the total number of observations is not necessarily an integer multiple of $s$, some observations may remain unused. To incorporate these observations, the segmentation procedure is repeated from the opposite end of the profile. Consequently, a total of $2N_s$ segments are obtained.

A local trend $y_v(i)$ of order m is fitted to each segment $v$, where the variance in the segment is then calculated as follows:

$$F^2(s,v) = (1/s)\sum_{i=1}^{s} \left\{Y[(v-1)s+1] - y_v(i)\right\}^2, \; v = 1, \ldots, N_s \tag{3}$$

Also

$$F^2(s,v) = (1/s)\sum_{i=1}^{s} \left\{Y[N-(v-1)s+1] - y_v(i)\right\}^2, \; v = N_s = 1, \ldots, 2N_s. \tag{4}$$

For segments obtained from the opposite end of the series, the variance is calculated analogously. The variances obtained from all $2N_s$ segments are subsequently combined to calculate the generalized fluctuation function.

For $q \neq 0$, the generalized fluctuation function is defined as:

$$F_q(s) = (1/2N_s)\sum_{v=1}^{2N_s} \left\{F^2[(s,v)]^{q/2}\right\}^{1/q} \tag{5}$$

The parameter $q$ represents the statistical moment used to characterize the fluctuations and can assume positive or negative real values. Positive values of $q$ assign greater weight to segments with large fluctuations, whereas negative values emphasize segments with small fluctuations. Thus, varying $q$ allows investigation of the scaling properties associated with different fluctuation magnitudes. In particular, $q$ =2 recovers the conventional DFA fluctuation function. In addition, for $q = 0$, the expression obtained via the mathematical limit as $q \to 0$ is used:

$$F_0(s) = \exp\left\{\left(1/4N_s\right)\sum_{v=1}^{2N_s} \ln\left[F^2(s,v)\right]\right\} \tag{6}$$

The scaling behavior of the fluctuations is examined by evaluating the dependence of $F_{q(s)}$ on the scale $s$. In the region where scaling behavior exists, a power-law relationship emerges as written:

$$F_q(s) = s^{H(q)} \tag{7}$$

The exponent $H(q)$, known as the generalized Hurst exponent, is estimated from the slope of the linear regression of log $F_q(s)$ versus log $s$:

$$H(q) = \frac{d\log\left[F_q(s)\right]}{d\log(s)} \tag{8}$$

In this formulation, $H(q)$ provides an important criterion for characterizing the scaling structure of the time series. For monofractal signals, the generalized Hurst exponent is approximately independent of $q$, but a direct dependence of $H(q)$ on $q$ implies multifractal behavior. Additionally, at $q$=2, the original Hurst exponent is recovered. For $H = 0.5$, the system is described as a random walk; if the result lies within the range $0 < H < 0.5$, the series is classified as long-range anti-correlated; and if *0.5 < H < 1,* it implies a positively long-range correlated series. From *H(q)*, it is possible to obtain the scaling exponent function *τ(q)* calculated as follows:

$$\tau(q) = qH(q) - 1 \tag{9}$$

The multifractal properties can also be characterized by the singularity spectrum *f(α)*, which describes the fractal dimension associated with different local scaling exponents. The Hölder exponent *α* is obtained via the Legendre transform:

$$\alpha = \frac{d\tau}{dq}, \tag{10}$$

and the corresponding fractal dimension is given by:

$$f(\alpha) = q\alpha - \tau(q)$$

(11)

It is also useful to obtain the width of the singularity spectrum, which is defined as:

$$\Delta\alpha = \alpha_{\max} - \alpha_{\min} \tag{12}$$

The width *Δα* measures the complexity of the fractality. The greater the difference, more complex the system is and the greater the multifractal richness observed.

### 2.3 Small-Shuffle Surrogate (SSS)

The Small-Shuffle Surrogate (SSS) method was originally proposed and described in detail by Nakamura and Small [36]. Unlike conventional random shuffling methods, which completely break the dataset temporal ordering, SSS preserves the time series embedded nonstationarity. As a result, it produces a surrogate series that disrupts local temporal organization while approximately preserving the original data overall structure and distribution. Because of these features, the SSS methodology is well-suited for investigating MEPP time series, biological signals known to exhibit nonstationary behavior.

## 3. Results

In this study, we applied the MFDFA procedure to the MEPP interval series using moment orders from $q$ = -20 to $q$ = 20, scales from $s$ = 10 to $s$ = 100, and a second-order polynomial detrending procedure (m = 2). We used these parameters for both the original and surrogate time series to compare their multifractal scaling properties directly. Only electrophysiological recordings containing ≥ 1000 intervals were selected for this study. This criterion allowed us to collect 108770 MEPP intervals from 52 recordings obtained at six different $Ca^{2+}$ concentrations. This strategy provided sufficient data for a rigorous multifractal analysis. Table 1 summarizes the descriptive statistics for the electrophysiological recordings used in our calculations. The results in Fig. 2, taken at $[Ca^{2+}]_o$ = 1.8 mM, are based on calculations performed on the original and SSS data. Additional simulations for $[Ca^{2+}]_o$ < 1.8 mM (hypocalcemia) and $[Ca^{2+}]_o$ > 1.8 mM (hypercalcemia) are shown in the supplementary material (Figs. S1-S2).

**Table 1**

Summary of descriptive statistics relative to electrophysiological data collected at different $[Ca^{2+}]_o$ levels. Averages are expressed as the mean ± standard deviation, calculated within each $[Ca^{2+}]_o$ group. MinNIR: Minimum number of MEPP intervals recorded in a given dataset; MaxNIR: Maximum number of MEPP intervals recorded in a given dataset.

| Concentration (mM) | Number of recordings | Total number of intervals | MinNIR | MaxNIR | Average |
|---|---|---|---|---|---|
| 0.6 | 7 | 11707 | 1019 | 2471 | 1672.42±586.83 |
| 1.2 | 6 | 13971 | 1195 | 3468 | 2328.50±766.96 |
| 1.8 | 10 | 19414 | 1009 | 3060 | 1941.40±723.34 |
| 4.8 | 5 | 6131 | 1008 | 1864 | 1226.20±360.38 |
| 10 | 13 | 31150 | 1228 | 3724 | 2396.15±668.97 |
| 15 | 11 | 26327 | 1064 | 4123 | 2393.36±887.61 |

Figure 2 (top panels) shows the plots for $H_q$ based on electrophysiological recordings at $[Ca^{2+}]_o$ = 1.8 mM. Table 2 summarizes all results at this concentration and shows that, on average, the original and SSS simulations agree. Relative to the original dataset, a small difference was found in data 28 ($\Delta H(q)$ = 0.27), but a large distinction was documented in data 36 ($\Delta H(q)$ = 1.23). For SSS data, the smallest value was in data 35 ($\Delta H(q)$ = 0.44), while the largest was in data 36 ($\Delta H(q)$ = 1.32). Comparing mean values between simulations using original and SSS data showed no significant difference. We also found no relationship between $\Delta H(q)$ and the number of intervals in these data. In the supplementary material (Fig. S1-S2; Table S1-S5), the analyses for $[Ca^{2+}]_o$ < 1.8 mM and $[Ca^{2+}]_o$ > 1.8 mM are available, where the results present a similar statistical profile compared to Table 2.

**Table 2**

Summary of descriptive statistics for generalized Hurst exponent calculations for $[Ca^{2+}]_o$ = 1.8 mM. Results are expressed as the mean ± standard deviation

| Data | $H_{min}$ (original) | $H_{max}$ (original) | ΔH(q)(original) | $H_{min}$ (SSS) | $H_{max}$(SSS) | ΔH(q)(SSS) |
|---|---|---|---|---|---|---|
| 28 | 0.47 | 0.74 | 0.27 | 0.41 | 0.94 | 0.53 |
| 30 | 0.37 | 0.97 | 0.60 | 0.40 | 0.93 | 0.53 |
| 32 | 0.43 | 0.87 | 0.44 | 0.44 | 0.92 | 0.47 |
| 33 | 0.46 | 0.88 | 0.42 | 0.36 | 0.94 | 0.58 |
| 34 | 0.43 | 1.34 | 0.90 | 0.35 | 1.44 | 1.09 |
| 35 | 0.36 | 0.93 | 0.57 | 0.35 | 0.79 | 0.44 |
| 36 | 0.29 | 1.57 | 1.23 | 0.22 | 1.54 | 1.32 |
| 37 | 0.36 | 0.91 | 0.55 | 0.40 | 0.94 | 0.54 |
| 38 | 0.22 | 0.76 | 0.54 | 0.22 | 0.80 | 0.57 |
| 39 | 0.34 | 1.32 | 0.98 | 0.43 | 1.41 | 0.97 |
| | 0.38±0.08 | 1.00±0.28 | 0.61±0.29 | 0.35±0.08 | 1.03±0.27 | 0.67±0.31 |

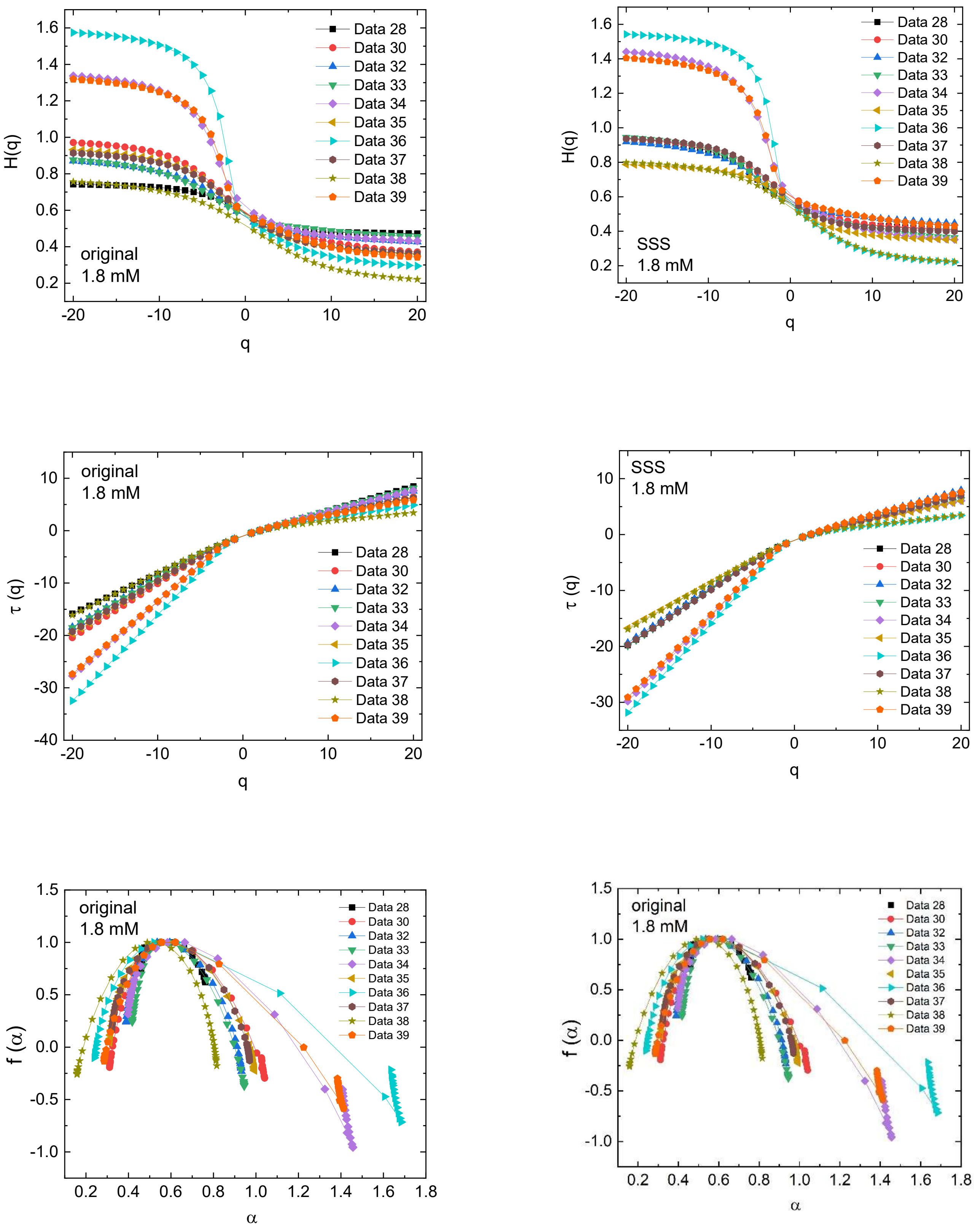

**Fig. 2** Summary, from top to bottom, of the numerical analysis of the generalized Hurst exponent, mass exponent, and multifractal spectra for $[Ca^{2+}]_o$=1.8 mM

Additionally, although the effect on the curve shape is less pronounced than for the generalized Hurst exponent, dispersion may also be observed relative to the mass exponent curves in all examined data for $[Ca^{2+}]_o$ = 1.8 mM. Nevertheless, for q > 0, the mass exponent is also more indistinguishable than for q > 0 in both the original and SSS data (Fig. 2, middle panel). Moreover, relative to the original dataset, the mass exponents for data 28 ($\Delta\tau(q)$ = 24.27) and 36 ($\Delta\tau(q)$ = 37.37) are the maximum and minimum, respectively. Furthermore, SSS analysis revealed data 38 ($\Delta\tau(q)$ = 20.45) and data 39 ($\Delta\tau(q)$ = 36.70) as the minimum and maximum mass exponents, respectively. Yet, for the original data, the descriptive statistical summary is (represented by mean ± S.D.): a) $<\Delta\tau(q)_{minimum}>$ = -21.59 ± 5.6; b) $<\Delta\tau(q)_{maximum}>$ = 6.48±1.54; c) $<\Delta\tau(q)_{range}>$ = 28.06 ± 5.50. In addition, for SSS data, one may mention the following values: a) $<\Delta\tau(q)_{minimum}>$ = -22.30 ± 5.64; b) $<\Delta\tau(q)_{maximum}>$ = 6.2 ± 1.58; c) $<\Delta\tau(q)_{range}>$ = 28.50 ± 5.58. In Fig. S3-S4, from the supplementary material, the analyses for $[Ca^{2+}]_o$ < 1.8 mM and $[Ca^{2+}]_o$ > 1.8 mM are displayed, where the curves also show the equivalent pattern of the physiological concentration values.

The multifractal spectrum curves for the original and SSS data show the characteristic multifractal shape for all 52 recordings (Fig. 2, lower panels). Table 3 provides descriptive statistics summarizing the width of Hölder exponents. In this case, one may note a significant variability in the exponents width mean values for the original data, spanning from 0.65 ± 0.13 ($[Ca^{2+}]_o$ = 0.6 mM) to 0.78 ± 0.30 for $[Ca^{2+}]_o$ = 1.2 mM) and 0.78 ± 0.37 ($[Ca^{2+}]_o$ = 15 mM, while for SSS the average lies between 0.60 ± 0.11 ($[Ca^{2+}]_o$ = 0.6 mM), and 0.83 ± 0.34 ($[Ca^{2+}]_o$ = 1.8 mM). Notably, the spectral width did not change significantly under hypocalcemic and hypercalcemic conditions compared with physiological concentrations. In this sense, the curve shape remains unaltered, indicating that the mechanisms supporting nerve-muscle communication are dynamically stable at the electrophysiological level. This observation holds for both analyses using the original data and those where SSS was performed. Undoubtedly, the richness of the spectrum is a signature of NMJ mechanisms adapting to extrinsic modifications imposed by the extracellular environment. Table 3 confirms that, despite fluctuations in the mean values and large deviations, results across concentrations do not differ significantly. Notably, we observed multifractal behavior regardless of the assumed $[Ca^{2+}]_o$.

**Table 3**

Summary of descriptive statistics for the multifractal spectrum width for each $[Ca^{2+}]_o$. Generalized Hurst exponent calculations for all $[Ca^{2+}]_o$. Results are expressed as the mean ± standard deviation

| $[Ca^{2+}]_o$ (mM) | Δα (original) | Δα (SSS) |
|---|---|---|
| 0.6 | 0.65±0.13 | 0.60±0.11 |
| 1.2 | 0.78±0.31 | 0.78±0.38 |
| 1.8 | 0.77±0.34 | 0.83±0.34 |
| 4.8 | 0.60±0.10 | 0.61±0.21 |
| 10 | 0.69±0.13 | 0.74±0.14 |
| 15 | 0.78±0.37 | 0.74±0.42 |

The results show that, although SSS destroys short-term variability (local structures) in the time series, the MFDFA results for both the original and SSS data remained similar, indicating that spontaneous electrical activity is governed by long-term temporal dynamics. Indeed, both the width of the multifractal spectrum and the structure of the generalized exponents are preserved in analyses of the original and SSS data. This provides particularly strong evidence for the robustness of the multifractal organization of spontaneous neuromuscular transmission, which is maintained for all 52 tissues analyzed at different calcium concentrations.

# 4. Discussion

Researchers have recognized and characterized fractal manifestations in the timing of electrical signals recorded from neurons across the nervous system [37-39]. Therefore, we do not doubt the utility of investigating the nature and physiological role of fractal concepts in the NMJ framework. Motivated by the importance of theme in neuroscience, we applied MFDFA to uncover a multifractal regime associated with communication between the nerve and diaphragm muscle in mice, based on the MEPPs time series. Our primary inspiration for revisiting this problem comes from other research groups that, over the years, have suggested a fractal nature behind the temporal dynamics of MEPPs, although conflicting results have also been reported. For example, Takeda *et al.* performed experiments in frog sartorius muscle and later analyzed event intervals using Higuchi method, providing evidence for fractality in MEPP occurrence [40]. Additionally, Lowen et al. found, from in vitro electrophysiological recordings at the Xenopus NMJ, that a fractal regime plays a role in the quantal release rate. [41]. However, the results of these studies were based on classical versions of the methodologies employed, only indicating the presence of monofractality under specific ionic conditions, and consequently establishing a more simplified physiological scenario. Furthermore, these studies did not address the role of fractality in pathological regimes.

In contrast, Kloot *et al.* used different theoretical approaches to examine MEPP intervals from the mouse diaphragm NMJ and the frog sartorius muscle, documenting a more complex framework [42]. These scholars found both the presence and absence of monofractality, where the heterogeneous results were attributed to a possible insensitivity of the methodology to the length of the time series. Therefore, developing theoretical approaches alongside rigorous experimental protocols remains necessary to identify and characterize fractal patterns in neurophysiological time series, particularly those related to neurotransmitter release.

The findings reported here offer a general theoretical methodology to uncover hidden processes underlying MEPP dynamics and acetylcholine release. A biophysical interpretation includes not only the temporal aspects of neurotransmission but may also involve spatial insights, since spontaneous discharge arises from synaptic vesicles anchored in active zones of the nerve terminal. The time-domain results in our measurements reflect spatial features driven by multiple mechanisms that interact synergistically to produce the multifractal persistence landscape shown in our results. Actin filaments in the presynaptic terminal could, even without electrical stimulation, contribute to an organized movement of several vesicles toward the active zones. This harmonic arrangement increases the likelihood of multiple docking, preventing vesicle traffic from following a random pathway [43-45]. Moreover, mature rodent NMJ synapses are characterized by active zones that contain few anchored vesicles, anatomically connected by protein structures [46]. In addition, exocytosis is widely accepted as a cyclic molecular machinery, typically described as a $Ca^{2+}$-dependent phenomenon. This means that when this ion remains temporarily elevated at the release site, long-range interactions governing the synaptic machinery strengthen, resulting in facilitation or attenuation of vesicular fusion at the same active zone [47,48]. Beyond that, spontaneous intracellular $Ca^{2+}$ oscillations also influence MEPP frequency [49]. Based on these arguments, the secretory machinery is a self-organized dynamic system despite perturbations induced by hypo- and hypercalemia.

The arguments above suggest that the cytosolic environment may behave as a system with spatial and temporal coherence, governed by a multifractal process acting on the molecular and structural apparatus of the exocytotic machinery. Therefore, given the $Ca^{2+}$ role in exocytosis, it would be important to perform electrophysiological recordings of spontaneous electrical activity while manipulating intracellular calcium concentration. This empirical maneuver would allow us to assess whether multifractality metrics change. Combining electrophysiological recordings with morphological measurements is another important goal for developing an integrated and deeper understanding of the multifractal role in neurotransmitter release. Finally, studies using mutant models that exhibit pathologies (degenerative diseases) affecting NMJ electrophysiological patterns are needed to expand the present results.

Reports show that the degree of multifractality depends on whether the physiological substrate is healthy or pathological. Investigators often applied these methods in *in vivo* studies comparing healthy controls with subjects with pathology. These studies claim that metrics derived from MFDFA calculations clearly distinguish between these two groups. For instance, Galaska *et al.* used this method in patients to identify patterns generated by healthy and pathologic states. Dutta *et al.* applied this analysis to human gait time series and showed that multifractality magnitude is more pronounced in the normal than in the diseased individuals [50,51]. Although such studies were conducted within specific age groups, it is known that aging can attenuate the fractal pattern in the brain [52]. For instance, *in vitro* reports indicate that neurons generate age-dependent electrical signals. Indeed, this has been confirmed in auditory system synapses during the maturation phase of their terminals [53]. Age modulations have been documented at the mouse NMJ, where MEPP frequency increased significantly with age [54]. Consequently, an important extension of the present work is to measure spontaneous electrical activity at the NMJ to assess whether the multifractality observed here is age-dependent. Also, electrophysiological recordings at the frog NMJ showed that frequency and amplitude are sensitive to $[Ca^{2+}]_o$ manipulation;

therefore, it would be appropriate to expand these findings by performing MFDFA calculations on MEPP amplitude as well [55, 56]. Using amplitudes for subsequent MFDFA analysis will provide important insights into the molecular processes underlying MEPP generation and the physiological role of the NMJ in the mammalian diaphragm.

Because we used only healthy adult animals, these profiles may reflect three characteristics of the diaphragm NMJ's, in preserving basal neurotransmitter release, despite ionic extracellular modulations: adaptability, flexibility, and robustness. These three elements may underlie this resilience, in which a synapse exhibits complex fluctuations with long-range correlations and self-affinity that govern the mechanisms responsible for preserving its physiological function. This explanation is supported by our previous work and other experimental studies in different NMJ preparations documenting populations of calcium-insensitive MEPPs [33,57]. From this resilience perspective, Saitoe *et al.* observed that removing spontaneous neurotransmitter release eliminates postsynaptic receptors, resulting in synaptic collapse [58]. Thus, the multifractal mechanism and, concomitantly, long-range correlations in MEPP time series drive metabolic adjustments within the healthy nerve terminal to preserve nerve-muscle communication under stress conditions, including hypo- and hypercalemic states.

Finally, the variability observed in the calculated metrics and curve representations may be, at least in part, attributable to experimental conditions, such as temperature fluctuations in the recording chamber. This empirical factor may consequently alter the basal resting membrane potential during the induction of the hypocalcemia and hypercalcemia protocols. Although our data were collected using a well-established *in vitro* protocol, in which tissues rested for an acceptable period before electrophysiological recordings began, residual mechanical trauma after tissue removal can still cause long-term injuries that may contribute to these fluctuations. Therefore, our criterion of using at least 1000 MEPP intervals in MFDFA calculations from a substantial number of tissues ensured reproducible and reliable results.

In summary, our findings strengthen the view that neurotransmitter release is strongly governed by a multifractal scenario, suitably represented by long-range correlations. Also, analyzing NMJ human tissues and rodent experimental models using MFDFA, may extend our main observations by providing a general physiological interpretation and clarifying the relationship between multifractality, plasticity mechanisms, and memory formation. Finally, we hope our contribution inspires the development of biophysical models that investigate the role of $Ca^{2+}$ in NMJ-related disorders.

# 5. Conclusion

Our results indicate that a complex organization is preserved across all six concentrations used to collect electrophysiological recordings. Remarkably, these findings suggest a robust biophysical mechanism in which changes in extracellular calcium were insufficient to alter the multifractal pattern. The exocytotic cycle operates according to multifractal rules that are robustly represented in the temporal dynamics of vesicular fusion and resist manipulation of extracellular calcium concentrations. Our results expand the scope of diaphragm NMJ applications as a genuine experimental model within complex systems. Using MFDFA on MEPP time series proved to be useful for identifying more general fractal structures that may not be adequately characterized by other simplified theoretical approaches. We hope this work advances understanding of the fractal properties of the electrical activity in neurotransmission and synaptic plasticity.

## CRediT authorship contribution statement

**A. J. da Silva**: Conceptualization, Supervision, Project administration, Investigation, Methodology, Visualization, Validation, Writing – review & editing, Writing – original draft, Formal Analysis. **S. Floquet**: Methodology, Visualization, Validation, Investigation, Formal Analysis, Software. **R.F. Lima**: Conceptualization, Methodology, Formal Analysis, Visualization, Validation, Investigation.

**Acknowledgments** Special thanks to Multi-User Facility of Drug Research and Development Center of Federal University of Ceará for technical support.

**Supplementary data** Supplementary material related to this article is also available.

## Declarations

**Conflict of Interest** The authors declare that there is no conflict of interest.

**Use of generative AI** We used ChatGPT for editing the section 2.2.

# Supplementary Information

# Persistence of multifractal dynamics at the mammalian neuromuscular junction across extracellular calcium conditions

**A. J. da Silva[1,*] ● S. Floquet[2] ● R.F Lima[3]**

[1] Centro de Formação em Ciências Agroflorestais, Universidade Federal do Sul da Bahia, Itabuna, Bahia, 45613-204, Brazil.

[2] Colegiado de Engenharia Civil, Universidade Federal do Vale do São Francisco, 48902-300, Juazeiro, Bahia, Brazil.

[3] Departamento de Fisiologia e Farmacologia, Faculdade de Medicina, Universidade Federal do Ceará, 60430-270, Fortaleza, Ceará, Brazil.

(*email: adjesbr@ufsb.edu.br)

Fig. S1: Results from the original data calculations for the generalized Hurst exponent under hypocalcemic ($[Ca^{2+}]_o < 1.8$ mM) and hypercalcemic ($[Ca^{2+}]_o > 1.8$ mM) conditions

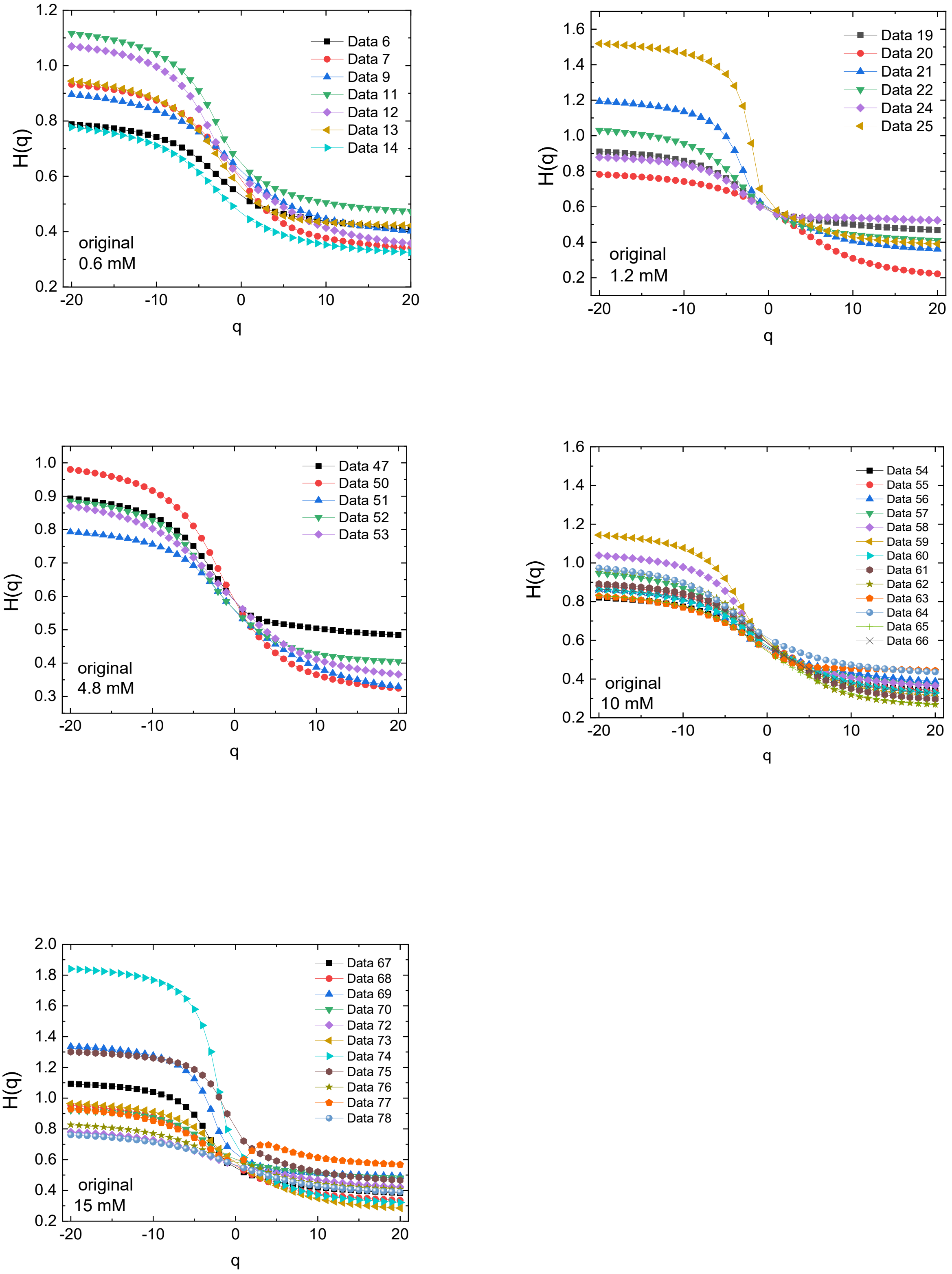

Fig. S2: Results of the SSS calculations for the generalized Hurst exponent under hypocalcemic ($[Ca^{2+}]_o$ < 1.8 mM) and hypercalcemic ($[Ca^{2+}]_o$ > 1.8 mM) conditions

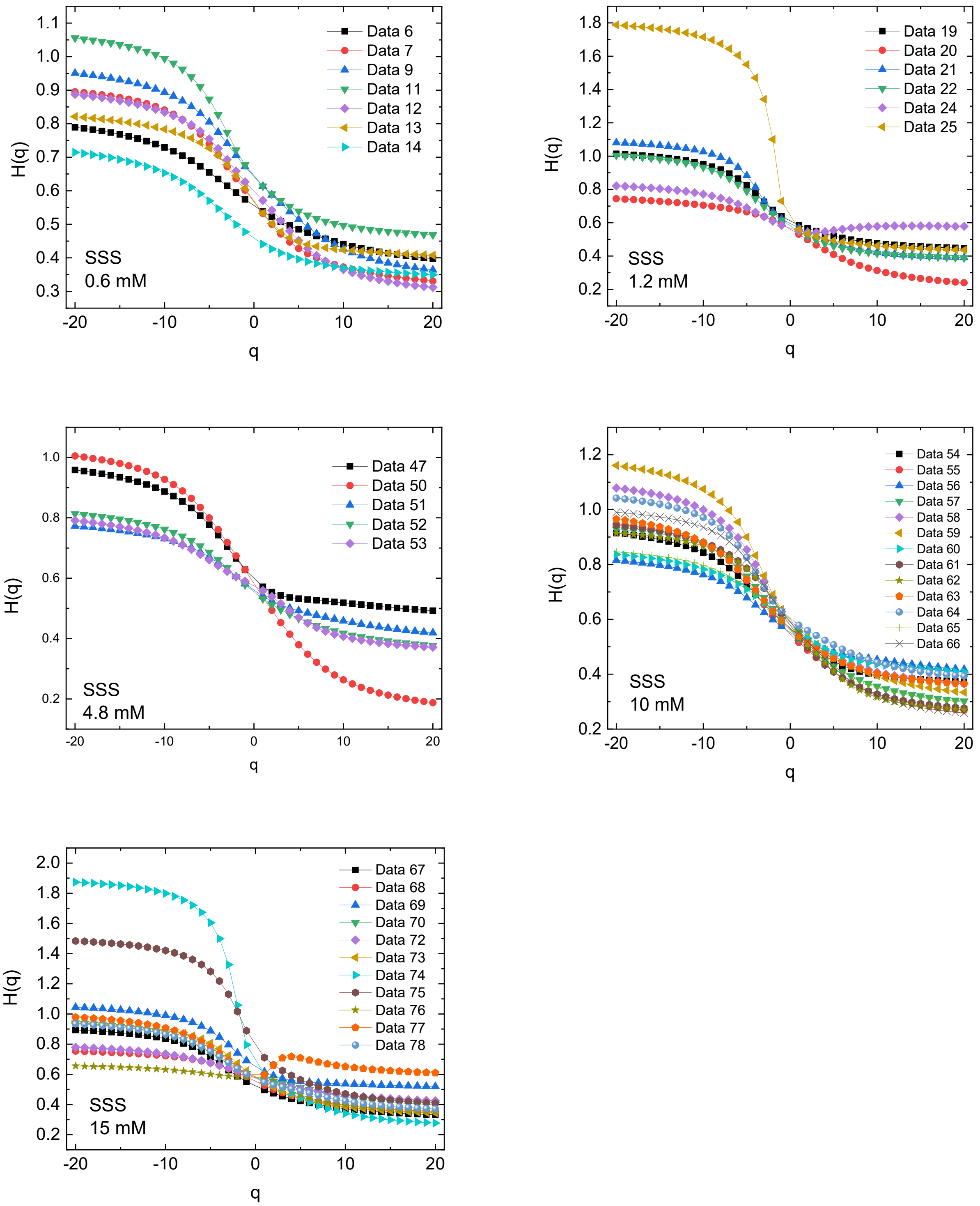

Fig. S3: Results of the original data calculations for the mass exponent under hypocalcemic ($[Ca^{2+}]_o < 1.8$ mM) and hypercalcemic ($[Ca^{2+}]_o > 1.8$ mM) conditions

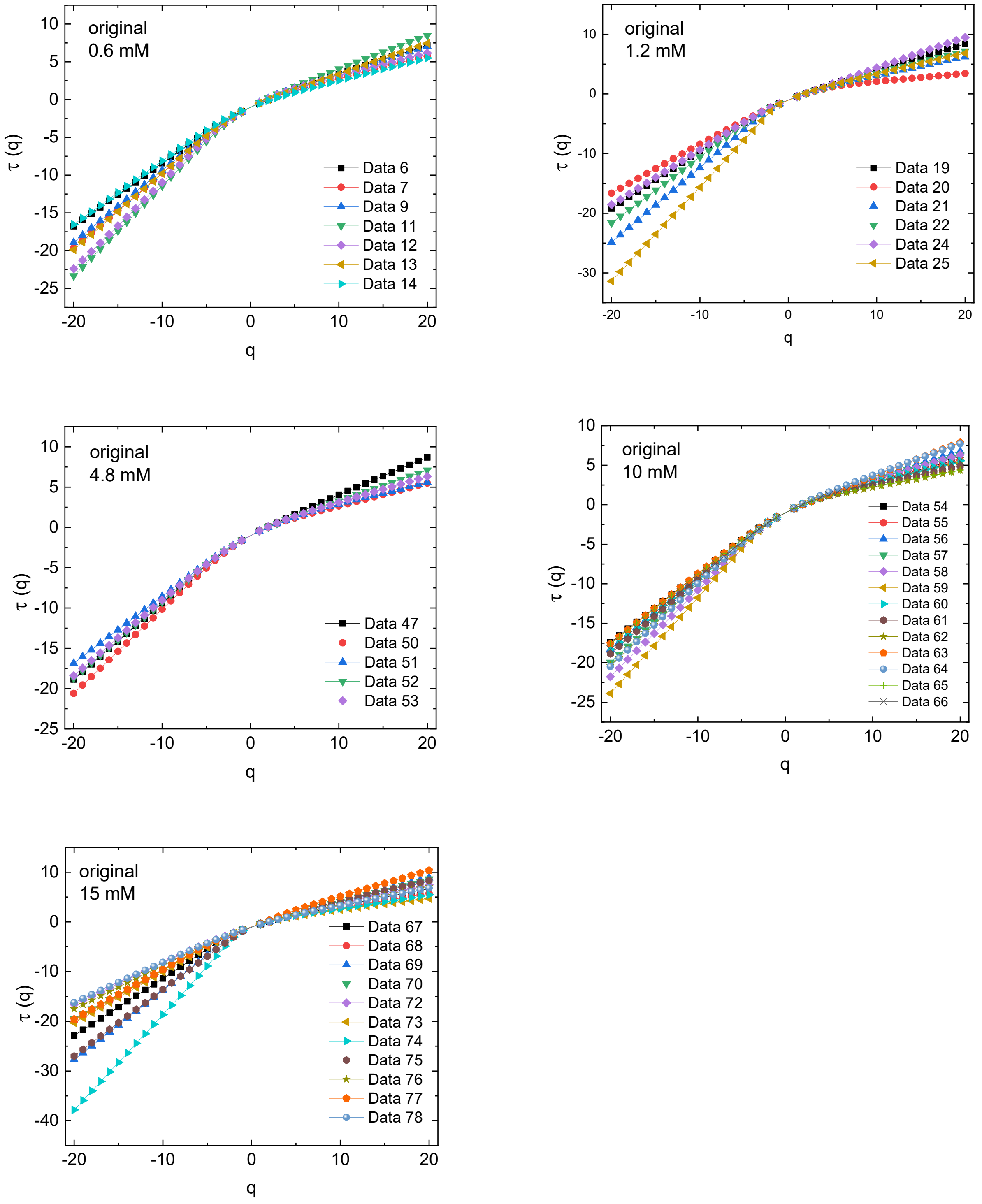

Fig. S4: Results of the SSS calculations for the mass exponent under hypocalcemic ($[Ca^{2+}]_o < 1.8$ mM) and hypercalcemic ($[Ca^{2+}]_o > 1.8$ mM) conditions

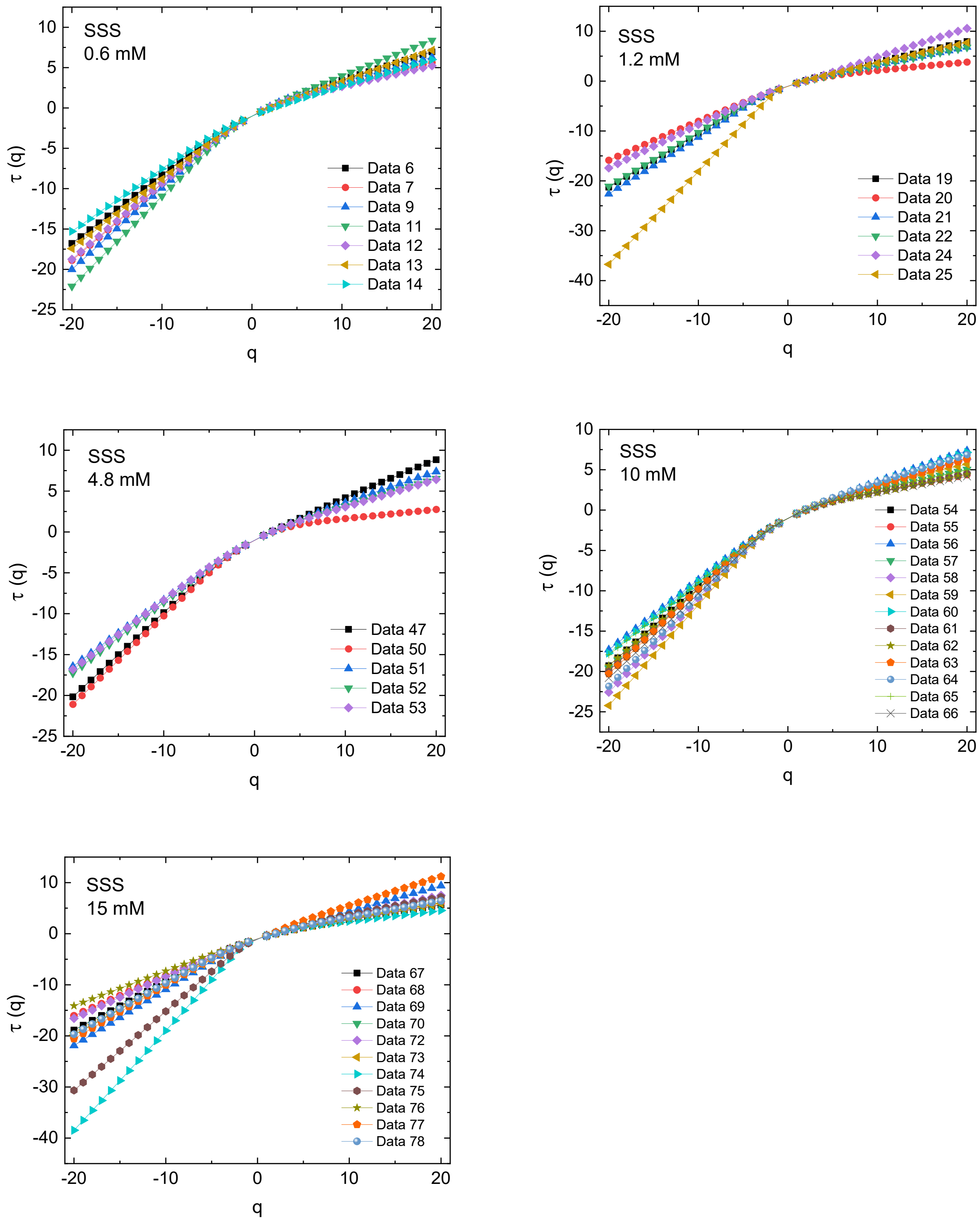

Fig. S5: Results of the original data calculations for the multifractal spectra under hypocalcemic ($[Ca^{2+}]_o < 1.8$ mM) and hypercalcemic ($[Ca^{2+}]_o > 1.8$ mM) conditions

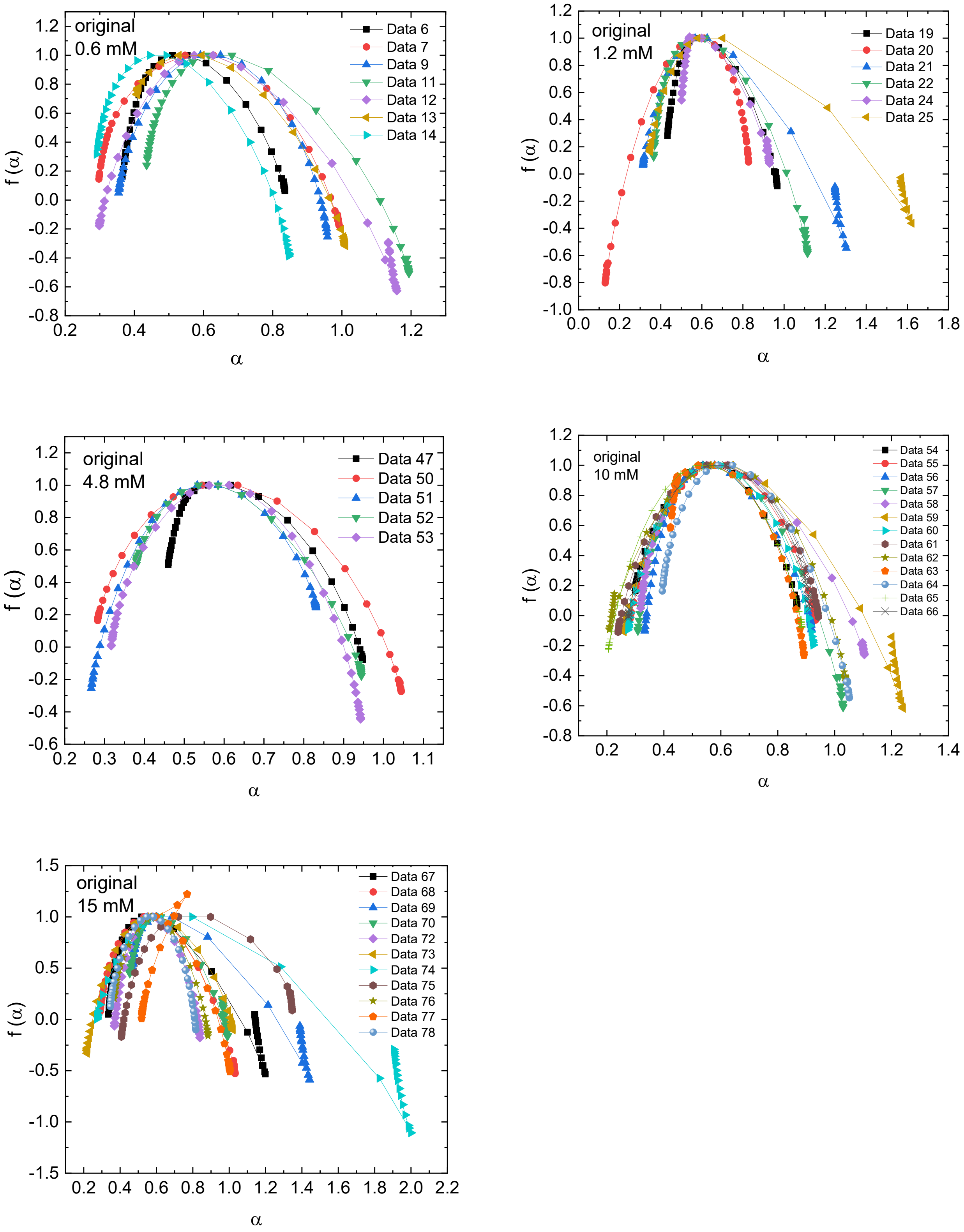

Fig. S6: Results of the SSS calculations for the multifractal spectra under hypocalcemic ($[Ca^{2+}]_o < 1.8$ mM) and hypercalcemic ($[Ca^{2+}]_o > 1.8$ mM) conditions.

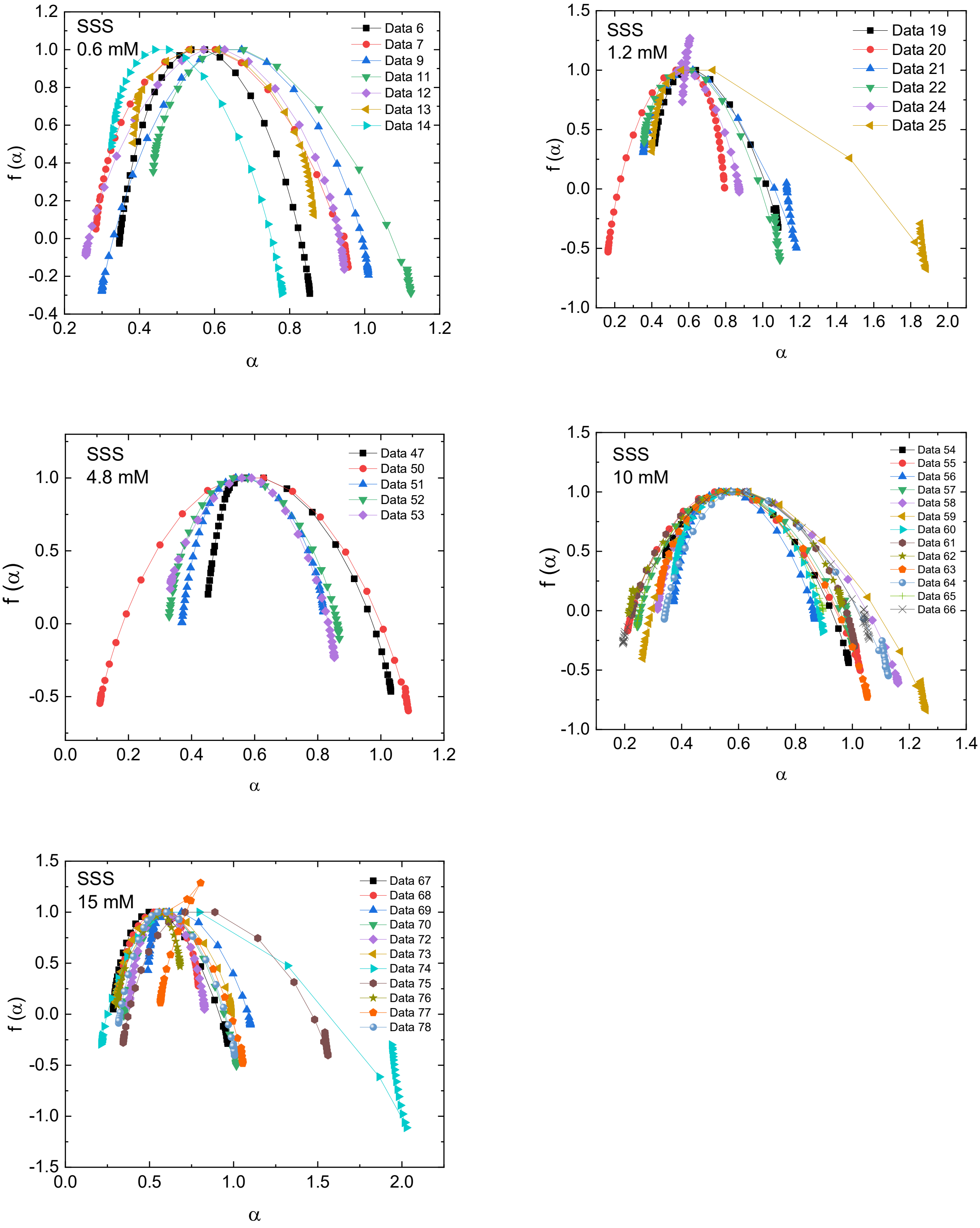

Table S1: Summary of descriptive statistics of the generalized Hurst exponent at $[Ca^{2+}]_o$ = 0.6 mM. Results are expressed as the mean ± standard deviation

| Data | $H_{min}$ (original) | $H_{max}$ (original) | ΔH(q)(original) | $H_{min}$ (SSS) | $H_{max}$(SSS) | ΔH(q)(SSS) |
|---|---|---|---|---|---|---|
| 6 | 0.41 | 0.79 | 0.38 | 0.40 | 0.79 | 0.39 |
| 7 | 0.34 | 0.93 | 0.59 | 0.33 | 0.90 | 0.56 |
| 9 | 0.40 | 0.90 | 0.49 | 0.36 | 0.95 | 0.59 |
| 11 | 0.47 | 1.12 | 0.64 | 0.47 | 1.06 | 0.59 |
| 12 | 0.36 | 1.07 | 0.71 | 0.31 | 0.89 | 0.58 |
| 13 | 0.42 | 0.94 | 0.52 | 0.41 | 0.82 | 0.41 |
| 14 | 0.32 | 0.78 | 0.45 | 0.35 | 0.71 | 0.37 |
| | 0.39±0.05 | 0.93±0.13 | 0.54±0.11 | 0.38±0.05 | 0.87±0.11 | 0.50±0.10 |

Table S2: Summary of descriptive statistics of the generalized Hurst exponent at $[Ca^{2+}]_o$ = 1.2 mM. Results are expressed as the mean ± standard deviation

| Data | $H_{min}$ (original) | $H_{max}$ (original) | ΔH(q)(original) | $H_{min}$ (SSS) | $H_{max}$(SSS) | ΔH(q)(SSS) |
|---|---|---|---|---|---|---|
| 19 | 0.47 | 0.91 | 0.44 | 0.45 | 1.01 | 0.57 |
| 20 | 0.22 | 0.78 | 0.56 | 0.24 | 0.74 | 0.50 |
| 21 | 0.36 | 1.19 | 0.83 | 0.39 | 1.08 | 0.69 |
| 22 | 0.41 | 1.03 | 0.62 | 0.40 | 1.00 | 0.62 |
| 24 | 0.52 | 0.88 | 0.36 | 0.54 | 0.82 | 0.28 |
| 25 | 0.39 | 1.52 | 1.13 | 0.44 | 1.79 | 1.35 |
| | 0.40±0.10 | 1.05±0.27 | 0.66±0.28 | 0.41±0.10 | 1.07±0.37 | 0.67±0.36 |

Table S3: Summary of descriptive statistics of the generalized Hurst exponent at $[Ca^{2+}]_o$ = 4.8 mM. Results are expressed as the mean ± standard deviation

| Data | $H_{min}$ (original) | $H_{max}$ (original) | ΔH(q)(original) | $H_{min}$ (SSS) | $H_{max}$(SSS) | ΔH(q)(SSS) |
|---|---|---|---|---|---|---|
| 47 | 0.48 | 0.89 | 0.41 | 0.49 | 0.96 | 0.47 |
| 50 | 0.32 | 0.98 | 0.66 | 0.19 | 1.00 | 0.82 |
| 51 | 0.33 | 0.79 | 0.46 | 0.42 | 0.77 | 0.35 |
| 52 | 0.41 | 0.89 | 0.48 | 0.38 | 0.81 | 0.44 |
| 53 | 0.37 | 0.87 | 0.50 | 0.37 | 0.79 | 0.42 |
| | 0.38±0.07 | 0.88±0.07 | 0.50±0.09 | 0.37±0.11 | 0.87±0.11 | 0.50±0.18 |

Table S4: Summary of descriptive statistics of the generalized Hurst exponent at $[Ca^{2+}]_o$ = 10 mM. Results are expressed as the mean ± standard deviation

| Data | $H_{min}$ (original) | $H_{max}$ (original) | ΔH(q)(original) | $H_{min}$ (SSS) | $H_{max}$(SSS) | ΔH(q)(SSS) |
|---|---|---|---|---|---|---|
| 54 | 0.34 | 0.82 | 0.48 | 0.37 | 0.91 | 0.54 |
| 55 | 0.33 | 0.88 | 0.55 | 0.27 | 0.95 | 0.68 |
| 56 | 0.38 | 0.86 | 0.48 | 0.42 | 0.85 | 0.40 |
| 57 | 0.36 | 0.95 | 0.58 | 0.30 | 0.93 | 0.63 |
| 58 | 0.37 | 1.04 | 0.67 | 0.37 | 1.08 | 0.71 |
| 59 | 0.32 | 1.14 | 0.82 | 0.33 | 1.16 | 0.83 |
| 60 | 0.33 | 0.86 | 0.54 | 0.41 | 0.84 | 0.43 |
| 61 | 0.30 | 0.89 | 0.60 | 0.28 | 0.94 | 0.67 |
| 62 | 0.27 | 0.97 | 0.70 | 0.27 | 0.91 | 0.64 |
| 63 | 0.44 | 0.82 | 0.39 | 0.37 | 0.97 | 0.60 |
| 64 | 0.44 | 0.97 | 0.54 | 0.39 | 1.04 | 0.65 |
| 65 | 0.27 | 0.83 | 0.56 | 0.30 | 0.85 | 0.55 |
| 66 | 0.32 | 0.86 | 0.54 | 0.26 | 0.99 | 0.73 |
| | 0.34±0.05 | 0.91±0.10 | 0.57±0.11 | 0.33±0.06 | 0.96±0.09 | 0.62±0.12 |

Table S5: Summary of descriptive statistics of the generalized Hurst exponent at $[Ca^{2+}]_o$ = 15 mM. Results are expressed as the mean ± standard deviation

| Data | $H_{min}$ (original) | $H_{max}$ (original) | ΔH(q)(original) | $H_{min}$ (SSS) | $H_{max}$(SSS) | ΔH(q)(SSS) |
|---|---|---|---|---|---|---|
| 67 | 0.38 | 1.09 | 0.71 | 0.33 | 0.89 | 0.56 |
| 68 | 0.34 | 0.95 | 0.62 | 0.36 | 0.75 | 0.40 |
| 69 | 0.49 | 1.34 | 0.84 | 0.52 | 1.04 | 0.53 |
| 70 | 0.48 | 0.92 | 0.44 | 0.40 | 0.94 | 0.54 |
| 72 | 0.42 | 0.78 | 0.36 | 0.42 | 0.78 | 0.36 |
| 73 | 0.28 | 0.96 | 0.68 | 0.35 | 0.94 | 0.60 |
| 74 | 0.32 | 1.84 | 1.52 | 0.28 | 1.87 | 1.60 |
| 75 | 0.47 | 1.30 | 0.83 | 0.41 | 1.48 | 1.08 |
| 76 | 0.41 | 0.83 | 0.42 | 0.35 | 0.66 | 0.30 |
| 77 | 0.57 | 0.93 | 0.36 | 0.60 | 0.98 | 0.38 |
| 78 | 0.39 | 0.76 | 0.37 | 0.37 | 0.93 | 0.56 |
| | 0.41±0.08 | 1.06±0.32 | 0.65±0.34 | 0.4±0.09 | 1.02±0.35 | 0.63±0.38 |